\documentclass[review,a4paper,3p,12pt]{elsarticle}
\journal{Chaos, Solitons \& Fractals}

\usepackage[T1]{fontenc}
\usepackage[utf8]{inputenc}
\usepackage{ams math,amssymb}
\usepackage[colorlinks=true,citecolor=ForestGreen,linkcolor=Red,urlcolor=Blue,hypertexnames=false]{hyperref}
\usepackage{chngcntr}

\usepackage{graphicx}

\usepackage{natbib}
\usepackage[dvipsnames]{xcolor}
\usepackage{hyperref}
\usepackage{booktabs}
\usepackage{color}
\counterwithout{figure}{section}

\begin{document}

\begin{frontmatter}
\title{Disagreement and fragmentation in growing groups}

\author[1,5]{Fanyuan Meng}
\author[1]{Jiadong Zhu}
\author[1]{Yuheng Yao}
\author[2]{Enrico Maria Fenoaltea}
\author[3]{Yubo Xie}
\author[4]{Pingle Yang}

\author[1,5]{Run-Ran Liu}
\author[1,5]{Jianlin Zhang}

\address[1]{Research Center for Complexity Sciences, Hangzhou Normal University, Hangzhou,311121, Zhejiang,China}

\address[2]{Physics Department, University of Fribourg, Chemin du Mus\'ee 3, 1700, Fribourg, Switzerland}
\address[3]{School of Computer and Communication Sciences,
    \'{E}cole Polytechnique F\'{e}d\'{e}rale de Lausanne, Lausanne, 1015, Vaud, Switzerland}

\address[4]{Business School, University of Shanghai for Science $\&$ Technology, 516 Jungong Rd., Shanghai 200093, China}

\address[5]{Corresponding authors: fanyuan.meng@hotmail.com; runranliu@163.com; jianlinzhang@hznu.edu.cn}

\begin{abstract}
The arise of disagreement is an emergent phenomenon that can be observed within a growing social group and, beyond a certain threshold, can lead to group fragmentation. To better understand how disagreement emerges, we introduce an analytically tractable model of group formation where individuals have multidimensional binary opinions and the group grows through a noisy homophily principle, i.e., like-minded individuals attract each other with exceptions occurring with some small probability.
Assuming that the level of disagreement is correlated with the number of different opinions coexisting within the group, we find analytically and numerically that in growing groups disagreement emerges spontaneously regardless of how small the noise in the system is. Moreover, for groups of infinite size, fragmentation is inevitable. We also show that the model outcomes are robust under different group growth mechanisms. 
\end{abstract}


\end{frontmatter}

\section{Introduction}

Social groups are collections of people who share similar interests \cite{tajfel2010social,stangor2015social}. They form spontaneously and continuously in our society \cite{newman2003social,jackson2010social}. Most human activities take place in groups, such as political parties, sports organizations, online communities, etc. Therefore, understanding their properties and formation is crucial to gain insights into social evolution and its practical implications.

Spontaneous group formation has been extensively studied in various disciplines such as sociology and psychology \cite{hogg1985interpersonal,backstrom2006group}. In the past two decades, physicists have also approached the problem of group formation and growth with quantitative tools borrowed from network science and complex systems \cite{barabasi1999emergence,cimini2019statistical, jusup2022social}. Mainstream research investigates how groups emerge in society \cite{girvan2002community,newman2004finding, minh2020effect, pham2021balance} and how the evolution and growth of a group affect its topological structure \cite{barabasi1999emergence,dorogovtsev2000structure,krapivsky2001degree, schweitzer2022social}. In these works, there is the implicit or explicit assumption that similar individuals attract each other (homophily assumption). Hence, groups, understood as clusters of individuals, emerge and stably grow through the aggregation of similar individuals \cite{mcpherson2001birds,kossinets2009origins, pham2022empirical}. However, real groups are not always stable. From the perspective of evolutionary game theory \cite{perc2013evolutionary}, the stability of a group following its formation and growth is determined by the costs and gains of individual actors, and of the group as a whole \cite{szolnoki2011group}. In this context, scholars investigate how the evolution of a group and the self-organization of its members leads to configurations that favor human cooperation (see \cite{perc2017statistical} for a recent review). Nonetheless, the agreement between members of a growing group is also determined by intrinsic factors such as the diversity or similarity of individuals (e.g., in terms of opinions), which are often independent of the members' pay-off to be part of the group. Actually, although cooperation and agreement are frequently correlated, they are different characteristics that follow different mechanisms.

There is evidence in the literature that the agreement between members is a rare exception \cite{acemoglu2010persistence,acemoglu2011opiniondyna,king2012disagreement,LIPIECKI2022112809}. Even if a group grows by the above-mentioned homophily principle, minor differences between individuals belonging to this growing cluster could be fatal enough to lead to the group's dissolution. Indeed, in the real world, it is not rare to see an initially stable and homogeneous social group (e.g., a small group of friends) fragmenting after its growth because of the arising of disagreement between individuals \cite{palla2007quantifying, bhowmick2020splitting, flamino2021creation}. This is also corroborated by experimental studies in sociology and psychology that suggest that group cohesion decreases with the group size \cite{delhey2007enlargements,soboroff2012group}. 
Nevertheless, the underlying mechanism of this phenomenon is not yet understood, and the emergence of disagreement and fragmentation within a growing group is still an understudied topic. Hence, the question we address in this paper is: when a group grows with the homophily principle, how does disagreement emerge within the group?

To this end, we propose a simple model of group formation in which individuals have binary opinions (e.g., like or dislike) on different topics. Our work is a multidimensional generalization of a recent model proposed by three of us in \cite{https://doi.org/10.48550/arxiv.2107.07324}, where individuals have a binary opinion on only one subject. Specifically, the model describes a small initial group of like-minded individuals that grow through the admission of new members. Since we are mainly interested in groups that form spontaneously, we assume that an individual, to be rejected or accepted in the group, must be evaluated by only one group member. This is not the case in organized groups, where applicants are admitted based on an aggregate decision of multiple members.
We formalize the homophily principle in the following way: the higher the overlap between the opinions of the evaluating member and the candidate, the higher the probability that the latter is accepted. Moreover, to take into account all exogenous and random factors of social interactions (such as hidden or partial information \cite{meng2022whom,fenoaltea2022local, fenoaltea2021stable}) we assume that evaluations are contrary to our homophily principle with some probability that we shall refer to as noise.
Differently from \cite{https://doi.org/10.48550/arxiv.2107.07324}, where the overlap between the opinions of two individuals is either full or zero, in the following model the multiplicity of topics on which people have an opinion results in a larger spectrum of possible overlaps between two individuals. As we shall show below, this leads to richer results in terms of implications.

In this framework, we measure the level of disagreement in the group in terms of the number of different opinions and the relative amounts of individuals adopting those opinions. In particular, if all individuals hold the same opinions on all topics, then the disagreement is zero; at the other extreme, where all possible combinations of binary opinions are present and adopted by the same proportion of people, then the disagreement is maximum and the group is fragmented.
Our goal is to study how the level of disagreement varies with the group size and the noise. We show analytically and numerically that disagreement is inevitable and, for large groups, fragmentation is the only possible outcome.

The rest of the paper is organized as follows. In Section \ref{sec:model} we define the group growth model. Its analytical investigation and discussion are treated in Section \ref{sec:result}, divided into three subsections, dealing with different opinion dimensions and growing mechanisms. Section \ref{sec:discussion} summarizes the results. In Section \ref{sec:Conclusion} we present some possible extensions and perspectives.

\section{The Model}\label{sec:model}
\begin{figure*}[htbp]
    \centering
    \includegraphics[width=0.8\linewidth]{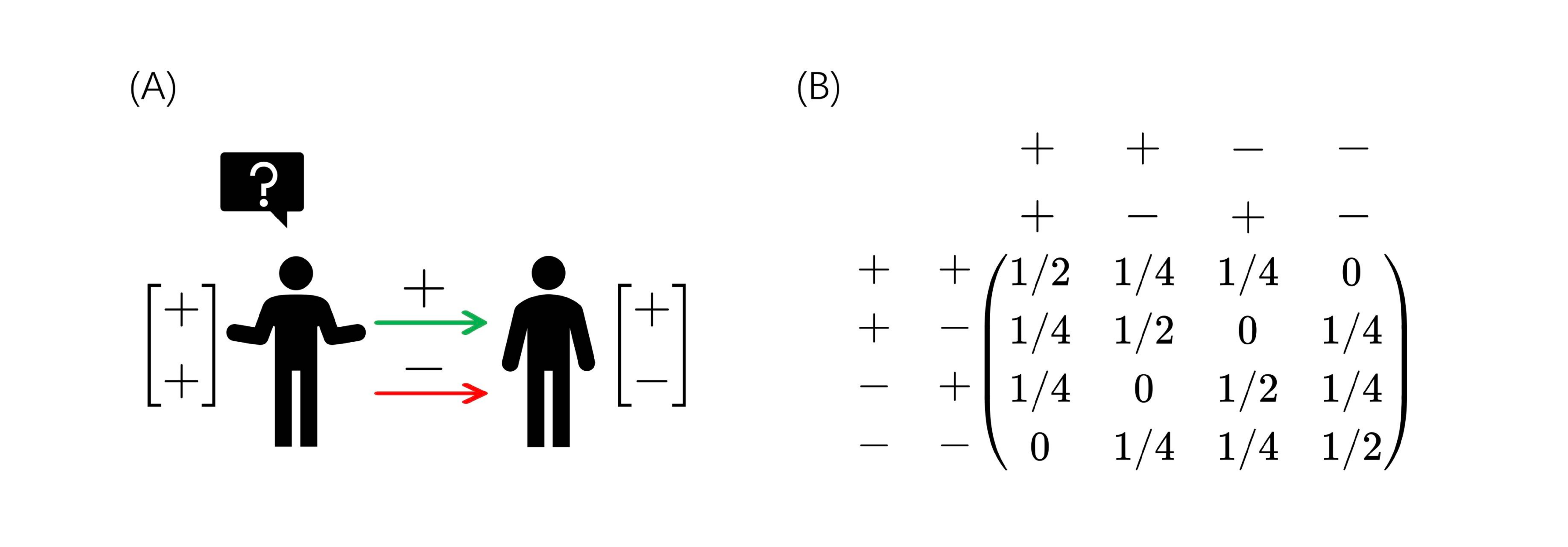}
    \caption{Illustration of the evaluation process and the acceptance probability matrix. (A) A candidate with opinions $(+1,-1)$ is evaluated by a group member with opinions $(+1,+1)$. Thus, the opinion vector overlap is zero, and the probability that the candidate is accepted is 1/2. (B) The acceptance probability matrix for $d=2$.}
    \label{fig:model_illu}
\end{figure*}

We assume that individuals have positive or negative opinions on $d$ different subjects. Each individual's opinion $o$ is formally represented by a $d$-dimensional vector whose elements can be either $+1$ (positive opinion) or $-1$ (negative opinion). Thus, there are $2^d$ different possible opinion vectors in the opinion vector space. The group growth is defined as follows. Initially, the group is composed of $N_0$ members with the same opinion vector, which we assumed to be $o_1=\{+1\}^d$, i.e., $d$ positive opinions. In each subsequent step, a new individual $y$ with opinion vector $o_y$ must be rejected or accepted in the group (see Fig.\ref{fig:model_illu}A). The sign of each $o_y$ element is drawn at random with equal probability. Thus, the opinion vector $o_y$ of each new individual is randomly chosen from the opinion vector space. The individual $y$ is then evaluated by a member $x$ of the group with opinion vector $o_x$. Based on the homophily principle that similar people like each other, we assume that the acceptance probability $M_{xy}$ that a member $x$ accepts in the group an individual $y$ is  
\begin{equation}\label{eq:admission_prob}
M_{xy}=\frac{1}{2}+\frac{K_{xy}}{d}\left(\frac{1}{2}-\eta\right).
\end{equation}
Here, $K_{xy}=o_x \cdot o_y$ is the scalar product of the two opinion vectors, and $\eta \in [0,0.5]$ is a control parameter that characterizes the level of noise in the evaluation. In particular, when $\eta=0$, the acceptance probability fully reflects the opinion overlap between $x$ and $y$: if $o_x=o_y$ ($x$ and $y$ agree on every subject), then $K_{xy}=d$ and $M_{xy}=1$, meaning that $y$ will be surely accepted by $x$. If $o_x=-o_y$ ($x$ and $y$ disagree on every subject), then $K_{xy}=-d$ and $M_{xy}=0$, meaning that $y$ will be surely rejected by $x$. In the intermediate case where the two individuals agree only on half of the subject, then $K_{xy}=0$ and $M_{xy}=1/2$, meaning that $y$ will be accepted or rejected with equal probability.
On the other hand, when $\eta=1/2$, $M_{xy}=1/2$ independently of the opinion vector overlap.
The group growth continues until the desired size of $N$ group members is reached.

In the following, we consider two ways of choosing the evaluating member $x$ from the group: 1) $x$ is chosen uniformly at random (uniform choice, UC); 2) $x$ is a member chosen with a preferential attachment probability (PA), i.e., with a probability proportional to the positive evaluations already done. While the UC is analytically easier, the PA is more realistic as it involves with a higher probability the most popular group members, as usually occurs in real social systems \cite{barabasi1999emergence,papadopoulos2012popularity}.

Defining $\overline{p_x}:=N_{x}/N$ as the fraction of members within the group having the same opinion vector $o_x$, our goal is to study how $\overline{p_x}$ varies as a function of noise $\eta$ and group size $N$. Note that, if $\overline{p_x}=1$ for a certain $o_x$, then the group has a uniform opinion, meaning that the level of disagreement is zero. On the other hand, if $\overline{p_x}=1/{2^d}$ for any $x$, the group has the highest level of disagreement as all opinion vectors coexist with the same weight: the system is said to be fragmented.

\section{Results}\label{sec:result}

\subsection{Preliminaries: the case of 1-dimensional opinion ($d=1$)}
The simplest (but far from reality) system to study is the one where individuals have a binary opinion on only one subject. This corresponds to having opinion vectors with $d=1$. In this case, the model is identical to the original model in \cite{https://doi.org/10.48550/arxiv.2107.07324}, which therefore serves as the base for this paper. In \cite{https://doi.org/10.48550/arxiv.2107.07324}, the fraction of members with positive opinions is interpreted as a measure of group cohesion. With our formalism, group cohesion is $\overline{p_{1}}$ with $o_{1}=\{+1\}$. Nevertheless, when we allow opinions to have more than one dimension, the most suitable interpretation of $\overline{p_x}$ is related to the level of agreement or disagreement in the group (note, however, that in common language the concepts of cohesion and agreement are somehow interdependent). Hence, to unify the language, we shall allude to disagreement and fragmentation also when $d=1$.

In this case, $M_{xy}=1-\eta$ if $o_x$ and $o_y$ have the same sign, and $M_{xy}=\eta$ if they have opposite sign. In other words, individual $x$ accepts individual $y$ with probability $1-\eta$ if they have the same opinion, and with probability $\eta$ if they have opposite opinions.

Following the arguments in \cite{https://doi.org/10.48550/arxiv.2107.07324}, it is possible to show that, for $N\gg N_0$, $\overline{p_{1}}$ has the following behavior:

\begin{equation}\label{eq:d1}
    \overline{p_{1}} \sim \left\{
    \begin{aligned}
    &\frac{1}{2}+ \frac{\Gamma(N_0+1)}{2\Gamma(N_0+1-2\eta)}N^{-2\eta} \quad \text{(UC)}, \\
    &\\
    &\frac{1}{2}+ \frac{(1-2\eta)\Gamma(N_0(N_0-1)/2+1)}{2(1-\eta)\Gamma(N_0(N_0-1)/2+1-\eta)}N^{-\eta}\quad \text{(PA)}.
    \end{aligned}
    \right.
\end{equation}

Clearly, the fraction of individuals with negative opinions are equal to $\overline{p_2}=1-\overline{p_{1}}$. Eq.\ref{eq:d1} shows that, in both UC and PA cases, the fraction of members with $o_x=\{+1\}$ in a system of size $N$ decreases rapidly with the noise $\eta$. Moreover, $\overline{p_1} \to 1/2$ for $N\to \infty$, even for infinitesimal levels of noise. This is shown in Fig.\ref{fig:UC}A and Fig.\ref{fig:UC}E for the UC case and in Fig.\ref{fig:PA}A and Fig.\ref{fig:PA}E for the PA case. These results imply that, even for small $\eta$, there is always a non-negligible degree of disagreement and, most importantly, when the group size goes to infinite the level of disagreement tends to its maximum regardless of how small is the noise. Hence, fragmentation is inevitable in large groups.

In the rest of this paper, we will generalize this result to the case of multidimensional opinions. In particular, we will show how the UC case and the PA case (in any dimension) can be unified through a mathematical formalism based on the spectral properties of the so-called acceptance probability matrix, which contains information about the outcome of an evaluation for every possible pair of opinion vectors.

\subsection{Analytical results for the UC case in $d$ dimension}\label{UC}

We first start with the simplest case in which, in each step, the evaluating member is randomly chosen (UC).

To obtain a general analytical result in the $d$ dimension, let us define the time of the process by the number of members accepted in the group. Thus, the initial members are in the group at time $t=0$, the first admitted individual, i.e., the $(N_0+1)$-th member, enters at time $t=1$, and so on. Finally, the $N$-th member enters at time $t=N-N_0$. Let us denote by $P_y(i)$ the probability that the individual admitted at time $t=i$ has an opinion vector $o_y$. We want to compute first $P_y(t)$ and then use it to compute $\overline{p_y}$ as 

\begin{equation}\label{eq:mean}
\overline{p_y}=\frac{1}{N}\left(N_0 \delta_{o_yo_1}+ \sum_{t=1}^{N-N_0}P_y(t)\right),    
\end{equation}
where $\delta$ is the Kronecker delta that is 1 for $o_y=o_1$ and 0 otherwise. Now, the probability $W_y(i)$ that an individual with opinion $o_y$ is accepted by member $i$ reads

\begin{equation}
    W_y(i) \propto \sum_{x=1}^{2^d} M_{yx} P_x(i).
\end{equation}
Since the time of the process counts only accepted members, once the evaluating member $i$ is chosen we want $\sum_yW_y(i)=1$ at each time-step. With this normalization constraint, we can write
\begin{equation}
    W_y(i) = \sum_{x=1}^{2^d} \frac{M_{yx}}{2^{d-1}} P_x(i).
\end{equation}

Since the individual entering the group at time $t+1$ is evaluated by a random member $i$ with $i \le t$, the probability $P_y(t+1)$ is given by

\begin{equation} \label{eq:pre_total}
    P_y(t+1) = \frac{N_0}{t+N_0}W_y(0) + \frac{1}{t+N_0}\sum_{\tau=1}^t W_y(\tau), 
\end{equation}

with the initial condition $P_y(0)=\delta_{o_yo_1}$. In Eq.\ref{eq:pre_total}, the first term on the right-hand side is the probability that the $t+1$-th member is evaluated and accepted by one of the initial members (entered the group at $t=0$). The second term is the probability that the $t+1$-th member is evaluated and accepted by one of the other $t$ members. With some algebraic manipulation, Eq.\ref{eq:pre_total} can be rearranged as:

\begin{equation}\label{eq:pre_form}
    (N_0+t)P_y{(t+1)}-(N_0+t-1)P_y{(t)}=\sum_{x=1}^{2^d} \frac{M_{yx}}{2^{d-1}} P_x(t).
\end{equation}

This equation relates the probability of having an opinion vector $o_y$ with the probabilities for all other opinion vectors in the opinion vector space. To solve it, we must rewrite it in a closed form where the probabilities of each opinion vector are interrelated. To this end, we define a vector $P(t)$ such that $P(t)=\{P_1(t),P_2(t),\dots,P_{2^d}(t)\}$ and define the acceptance probability matrix $M$ as a $2^d \times 2^d$ symmetric matrix in which each element $\{yx\}$ is given by $M_{yx}/2^{d-1}$. For example, when $d=1$, we have $M=\left(\begin{smallmatrix}
    1-\eta & \eta \\
    \eta & 1-\eta
\end{smallmatrix}\right)$ (for $d=2$, see Fig.\ref{fig:model_illu}B).

With this formalism, we can write the matrix form of Eq.\ref{eq:pre_form} as

\begin{equation}\label{eq:mat_form}
    (N_0+t)P{(t+1)}-(N_0+t-1)P{(t)}=MP(t),
\end{equation}

with the initial condition $P(0)=\{1,0,\dots,0\}$. By defining $B(t)\equiv [(1-\alpha(t))I+\alpha(t)M]$, where $I$ is the identity matrix and $\alpha(t)=1/(t+N_0)$, Eq.\ref{eq:mat_form} becomes
\begin{equation} \label{eq:mat_final}
    P(t+1)=B(t)P(t).
\end{equation}

Since $M$ is a real symmetric matrix, it can be decomposed as $M=Q\Lambda Q^T$, where $\Lambda$ is a diagonal matrix whose elements $\{\lambda_1, \lambda_2,\dots, \lambda_{2^d}\}$ are the eigenvalues of $M$ in descending order, while $Q$ is a matrix whose columns are the corresponding eigenvectors $\{q_1, q_2,\dots, q_{2^d}\}$ of $M$. In this way, $B(t)$ can be written as $B(t) = Q[(1-\alpha(t)) I+\alpha(t) \Lambda]Q^T$. With this formalism, Eq.\ref{eq:mat_final} can be solved analytically to obtain
\begin{equation} \label{eq:D}
    P(t) = Q\Big[ \prod_{i=1}^{t-1}\Big((1-\alpha(i)) I+\alpha(i) \Lambda \Big) \Big] Q^T P(1),
\end{equation}
where $P(1)=Q\Lambda Q^T P(0)$. Now, let us generalize Eq.\ref{eq:mean} as
\begin{equation}
\overline{p}=\frac{1}{N}\left(N_0P(0)+\sum_{t=1}^{N-N_0}P(t)\right)
\end{equation}
where $\overline{p}=\{\overline{p_1}, \overline{p_2},\dots, \overline{p_{2^d}}\}$. This sum can be written explicitly as $\overline{p}=QDQ^TP(0)$, where $D$ is a $2^d \times 2^d$ diagonal matrix whose elements are
\begin{equation}
   D_{yy}=\frac{\Gamma(\lambda_y+N)\Gamma(1+N_0)}{\Gamma(N+1)\Gamma(\lambda_y+N_0)} \quad \text{for $y=1, 2, \dots, 2^d$}.
\end{equation}
Using the explicit definition of the matrix $Q$, we can rewrite $\overline{p}$ as
\begin{equation}\label{eq:lambda_temp}
    \overline{p} = \sum_{y=1}^{2^d} \frac{\Gamma(\lambda_y+N)\Gamma(1+N_0)}{\Gamma(N+1)\Gamma(\lambda_y+N_0)} q_y q_y^T P(0).
\end{equation}

Since, by definition, the matrix $M$ has all non-negative entries and the sum of each column vector is equal to 1, then it is a Markov matrix \cite{gagniuc2017markov} and its largest eigenvalue must be $\lambda_1=1$ with the corresponding eigenvector $q_1=\{1/\sqrt{2^d}, 1/\sqrt{2^d},\dots, 1/\sqrt{2^d}\}$. In this way, we can write the final result as
\begin{equation}\label{eq:final_sol_UC}
    \overline{p} = q_1 q_1^T P(0)+\sum_{i=2}^{2^d} \frac{f(\lambda_i)}{f(\lambda_1)} q_i q_i^T P(0),
\end{equation}
where 
\begin{equation}\label{eq:fUC}
    f(\lambda_i)=\frac{\Gamma(N+\lambda_i)}{\Gamma(N_0+\lambda_i)} \quad \text{for $i=1, 2,\dots, 2^d$}.
\end{equation}

\begin{figure*}[htbp]
        \centering
        \includegraphics[width=0.8\linewidth]{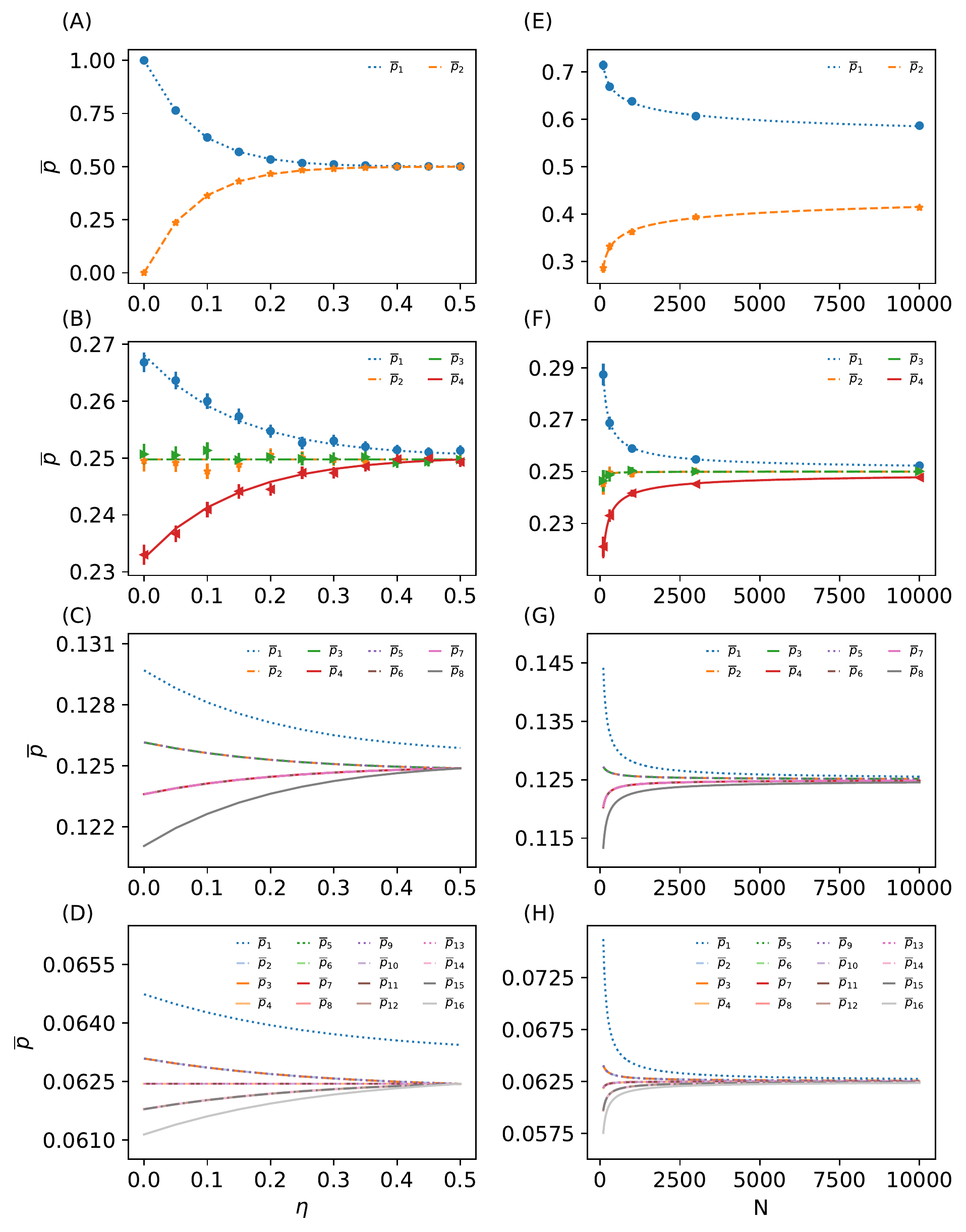}

        \caption{Averaged fraction $\overline{p_i}$ of group members with opinion $o_i$ in the UC case, for $i=1,2, \dots, 2^d$. (A-D) Relation between $\overline{p_i}$ and the level of noise $\eta$, with $d=\{1,2,3,4\}$ (A, B, C, D, respectively). Lines are the analytical solutions in Eq.\ref{eq:final_sol_UC} with $N=1000$ and $N_0=1$. Results show that, even when the noise is zero (excluding the case with $d=1$), there is no one dominant opinion vector, i.e., disagreement naturally emerges in the system. (E-H) Relation between $\overline{p}$ and the group size $N$, with $d=1,2,3,4$ (E, F, G, H, respectively). Lines are the analytical solutions in Eq.\ref{eq:final_PA} with $\eta=0.1$ and $N_0=1$. Here, plots show that, independently of the level of noise, the larger the group size the higher the disagreement, leading to full polarization/fragmentation in the extreme case of infinitely large systems.  In all panels, each point is the average of 1000 numerical simulations (the error bars show twice the standard error of the mean).}
        \label{fig:UC}
\end{figure*}

Notably, in the large $N$ limit, we have $\overline{p}\to \{1/2^d, 1/2^d,\dots, 1/2^d\}$ for any $\eta>0$, consistently with the $d=1$ case discussed in the previous section. The numerical results are in agreement with Eq.\ref{eq:final_sol_UC} and are shown in Fig.\ref{fig:UC}(A,B,E,F). We discuss the implications of these results in Section \ref{sec:discussion}. Before we briefly show the analytical procedure for the PA case.

\subsection{Analytical results for the PA case in $d$ dimension}\label{PA}
\begin{figure*}[htbp] 

        \centering
        \includegraphics[width=0.8\linewidth]{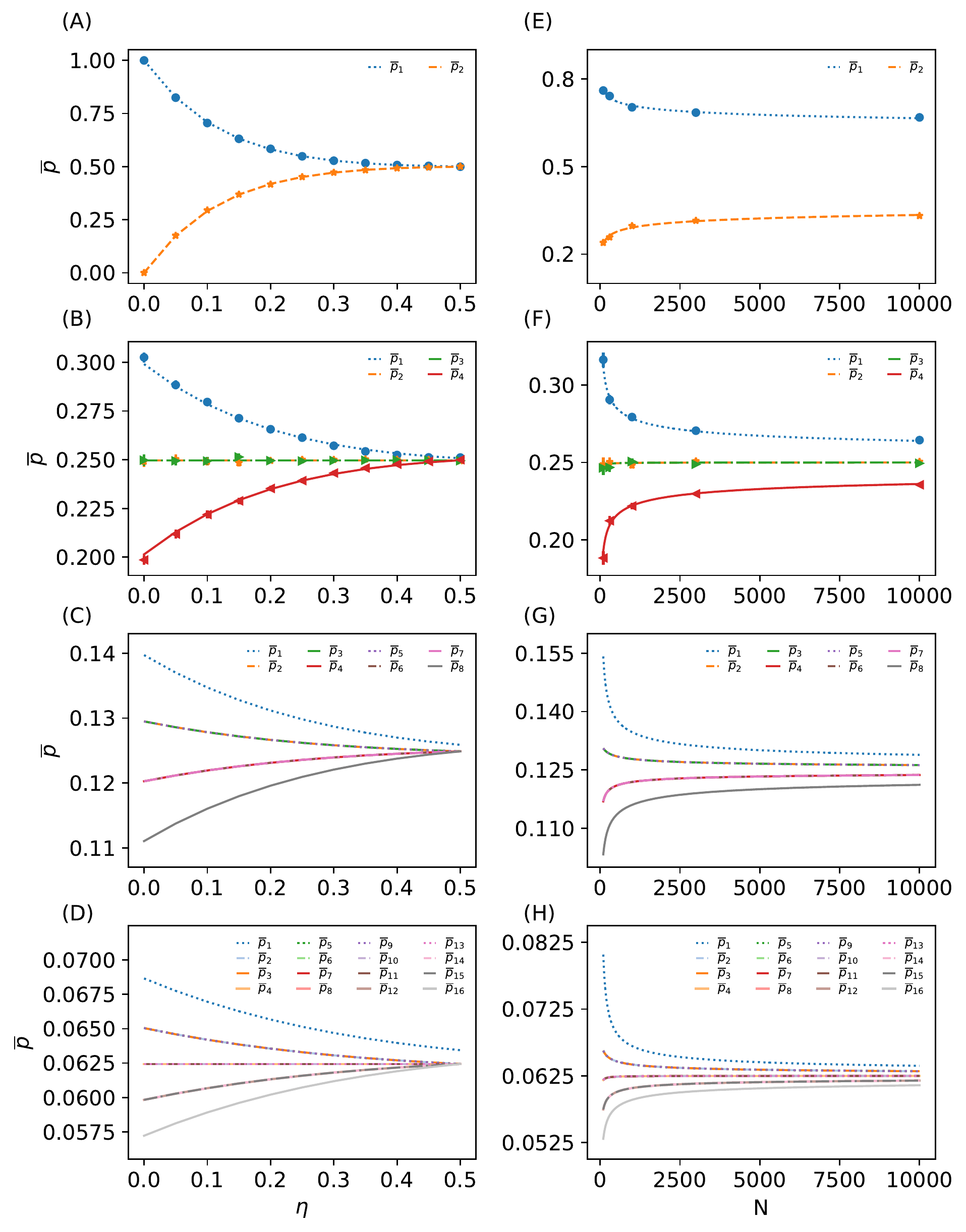}
        
        \caption{Averaged fraction $\overline{p_i}$ of group members with opinion $o_i$ in the PA case, for $i=1,2, \dots, 2^d$. (A-D) Relation between $\overline{p_i}$ and the level of noise $\eta$, with $d=\{1,2,3,4\}$ (A, B, C, D, respectively). Lines are the analytical solutions in Eq.\ref{eq:final_sol_UC} with $N=1000$ and $N_0=1$. (E-H) Relation between $\overline{p}$ and the group size $N$, with $d=1,2,3,4$ (E, F, G, H, respectively). Lines are the analytical solutions in Eq.\ref{eq:final_PA} with $\eta=0.1$ and $N_0=1$. The implications of these figures are the same as those described in Fig.\ref{fig:UC}. Here, however, the convergence to the full polarized state in the large $N$ limit is slower due to the PA mechanism. In all panels, each point is the average of 1000 numerical simulations (the error bars show twice the standard error of the mean).}
        \label{fig:PA}
\end{figure*}

In the PA case, the evaluating member is chosen at each time step with a probability proportional to the positive evaluations already done. The number of positive evaluations can be interpreted as a member's social capital or popularity. Formally, we denote by $k_i(t)$ the popularity at time $t$ of the member who entered the group at time $t=i$ (the time is defined as in Section \ref{UC}). We set the initial popularity $k_i(i)=1$ for each $i>0$; for the initial members, we set $k_0(0)=N_0-1$, since at $t=0$ they interact with all other $N_0-1$ initial members (for $N_0=1$, we set $k_0(0)=1$ to avoid the anomaly of having probability zero to choose an evaluating member). Following the prescription of Barabasi and Albert's model \cite{barabasi1999emergence}, the PA probability of being evaluated at time $t$ by the $i$-th member is $k_i(t)/(2t+N_0(N_0-1))$. The time evolution of the popularity of member $i$ is then given by
\begin{equation} \label{eq:k_evolution}
    k_i(t+1)=k_i(t) + \frac{k_i(t)}{2t+N_0(N_0-1)}. 
\end{equation}
Solving this recursive equation for the initial members and the other members separately (as there are two different initial conditions), we get
\begin{equation*}
    k_0(t) = \frac{1+2L}{N_0}\frac{g(t)}{g(0)},
\end{equation*}
\begin{equation}\label{eq:ki}
    k_i(t) = \frac{g(t)}{g(i)} \quad \text{for $i>0$},
\end{equation}
where $L:=N_0(N_0-1)/2$ and $g(t):=\Gamma(1/2+L+t)/\Gamma(L+t)$.
Having now the probability of choosing the evaluating member $i$ at time $t$, we can write an equation analogous to Eq.\ref{eq:pre_total} for the evolution of probability $P_y(t)$ in the PA case:
\begin{equation} \label{eq:pre_total_PA}
    P_y(t+1)=\frac{N_0 k_0(t)}{2t+2L}W_y(0) + \sum_{\tau=1}^t\frac{k_{\tau}(t)}{2t+2L}W_y(\tau).
\end{equation}
Following the same procedure of Section \ref{UC}, we can rewrite this equation in its matrix form $P(t+1)=B'(t)P(t)$, where $B'(t)=[1-\alpha'(t)I+\alpha'(t)M]$ with $\alpha'(t)=1/(2t+2L)$. Again, this equation can be decomposed and solved to obtain the final form of the vector $\overline{p}$ in the PA case:
\begin{equation}\label{eq:final_PA}
    \overline{p} = q_1 q_1^T P(0)+\sum_{i=2}^{2^d} \frac{f(\lambda_i)}{f(\lambda_1)} q_i q_i^T P(0),
\end{equation}
where 
\begin{equation}\label{eq:fPA}
    f(\lambda_i)=1+\frac{\lambda_i}{\lambda_i+1}(N_0-1)\Big(g(\lambda_i)-1\Big),
\end{equation}
and 
\begin{equation}
    g(\lambda_i)=\frac{\Gamma \Big(\dfrac{1+\lambda_i}{2}+N+\dfrac{1}{2}N_0(N_0-3)\Big)\Gamma(L)}{\Gamma\Big(N+\dfrac{1}{2}N_0(N_0-3)\Big)\Gamma\Big(\dfrac{1+\lambda_i}{2}+L\Big)},
\end{equation}
where $i=1, 2,\dots, 2^d$.
For $N\to \infty$, we still have $\overline{p}\to\{1/2^d, 1/2^d,\dots, 1/2^d\}$ for any $\eta>0$. These results are consistent with numerical simulations shown in Fig.\ref{fig:PA}(A,B,E,F).

\section{Discussion}\label{sec:discussion}
The results of Section \ref{UC} and \ref{PA} can be summarized as follows. Given a noise level $\eta$, the fraction of members with opinion $o_x$ in a group of $N$ individuals is given by Eq.\ref{eq:final_sol_UC} (Eq.\ref{eq:final_PA}) with  $f(\lambda_i)$ given by Eq.\ref{eq:fUC} in the UC case and by Eq.\ref{eq:fPA} in the PA case. Note that, with $d=1$, we restore the results of \cite{https://doi.org/10.48550/arxiv.2107.07324}. Indeed, for $d=1$ we have
$M=\left(\begin{smallmatrix}
    1-\eta & \eta \\
    \eta & 1-\eta
\end{smallmatrix}\right)$,
with eigenvalues $\{\lambda_1,\lambda_2\}=\{1,1-2\eta\}$ and eigenvectors $q_1=\{1/\sqrt{2},1/\sqrt{2}\}$ and $q_2=\{-1/\sqrt{2},1/\sqrt{2}\}$. Inserting these in Eqs.\ref{eq:final_sol_UC} and \ref{eq:final_PA}, we obtain Eq.\ref{eq:d1}.

The advantage of our approach is that we have constructed a unified formalism to represent the evolution of a group's opinions in any dimension and, most importantly, all information is incorporated into the matrix $M$ and its spectral properties.

The main implications of our results are the following. First, in both the UC and PA cases, even a small level of noise generates a consistent level of disagreement, i.e., the coexistence of different opinions within the group. Note that in the PA case, for a given level of noise, there is less disagreement, i.e., one opinion prevails more than the others. This is a consequence of the preferential attachment mechanism that makes older members (who entered the group earlier) more popular, thus favoring the acceptance of individuals more similar to them. Technically, this can be seen by comparing $B(t)$ with $B'(t)$ and noting that $\alpha(t)<\alpha'(t)$ for any $t>1$. This implies that when $P(0)=\{1, 0,\dots, 0\}$, $\overline{p_1}$ in the PA case is larger than that in the UC case (for a given $\eta$ and $N$).

Second, in both the UC and PA cases the stationary solution (i.e., for $N\to \infty$) of Eq.\ref{eq:final_sol_UC} is $\overline{p}=\{1/2^d, 1/2^d,\dots, 1/2^d\}$ independently of the noise level. Hence, in large groups, all opinions coexist with the same weight. Remarkably, this suggests that fragmentation is an inevitable phenomenon in sufficiently large social organizations. Hence, even if differences between members who enter the group early are infinitesimal (because the noise is also infinitesimal), they increasingly amplify in new members, until the group becomes fragmented.

Finally, note that opinion evolution is affected by the opinion dimension. Indeed, there is a difference between $d$ odd and $d$ even.  In the former case, the fraction of members with opinion $o_x$ is either always increasing or always decreasing in both $\eta$ and $N$, as shown in Fig.\ref{fig:UC} (A,E,C,G) and Fig.\ref{fig:PA} (A,E,C,G). However, when $d$ is even, there exists a fraction of individuals with opinion $o_{x'}$ that is constant and thus is not affected by either $\eta$ or $N$, as shown in Fig.\ref{fig:UC} (B,D,F,H) and Fig.\ref{fig:PA} (B,D,F,H). This happens if $K_{1x'}=o_1 \cdot o_{x'}=0$, that is, when there are elements in the matrix $M$ that do not depend on $\eta$. This condition is satisfied only when $d$ is even.

\section{Conclusion} \label{sec:Conclusion}
To summarize, we proposed a model of noisy group formation to study how disagreement and fragmentation emerge within growing social groups. Our framework complements and extends the original model proposed in \cite{https://doi.org/10.48550/arxiv.2107.07324}. In particular, we depart from the unrealistic assumption of Ising-like individuals (who can have only two opinions such as positive or negative), and introduce a mathematical methodology to study individuals with multidimensional opinions. In our framework, all the information about group evolution is defined by the spectral properties of the matrix of the acceptance probabilities.

Our findings suggest that, regardless of the amount of noise in the system, disagreement inevitably emerges as the group grows. Moreover, when the group grows infinitely, disagreement reaches its maximum level, i.e., the group is fragmented. In other words, fragmentation is inevitable in large social groups.  

Our model is an oversimplification of society. Although the random factors involved in the interaction between two individuals can be well represented by the noise parameter introduced in this paper, the emergence of disagreement is a complex phenomenon and is the result of many factors that cannot be easily identified. Nevertheless, our work focuses on a specific aspect of group formation and proposes a complementary explanation to existing ones on the emergence of social fragmentation.

Our model can be extended in several ways. For instance, one can study the case of time-varying opinion vectors either due to factors exogenous to the group (e.g., the strength of an external field representing, for instance, the impact of social media) or due to peer influence. The latter effect can be modeled by describing the group as a network where each individual is influenced by his or her neighbors. Also, one can include the possibility that unhappy members (for example the members surrounded by strongly different individuals) leave the group. These dynamics can be naturally implemented with an evolutionary game-theory approach. While here we focused on the evolution of member similarity by neglecting any type of cost-benefit analysis, the latter can be included, for example, as a parallel aspect in the admission (or expulsion) of group members. Another direction is to study the case of multiple evaluating members. In this way, the decision to admit a new individual would be the aggregate result of several evaluating members (e.g., by majority voting). In this paper we have considered only one evaluating member to describe the spontaneous growth of a group, however considering multiple evaluating members would allow one to study different mechanisms of group formation and admission processes. 

\section{Acknowledgement}
We thank Yu Liu for his helpful comments. 
\bibliographystyle{unsrt}
\bibliography{CSF.bib}

\end{document}